\documentclass[11pt]{article} 

\usepackage{amsmath}
\usepackage{amssymb}     
\usepackage{latexsym}

\newcommand{\dda}{\mathord{\mbox{\makebox[0pt][l]{\raisebox{-.4ex}
                           {$\downarrow$}}$\downarrow$}}}
\newcommand{\dua}{\mathord{\mbox{\makebox[0pt][l]{\raisebox{.4ex}
                           {$\uparrow$}}$\uparrow$}}}
\def\endproof{$\ \ \Box$} 

\newcommand{\bq}{\begin{quote}}
\newcommand{\eq}{\end{quote}} 

\newcommand{\convex}{{\mathbf C}}

\newcommand{\Pcom}{{\mathcal P}_{\mathit com}}

\newcommand{\EM}{\ll_{_{\mathrm EM}}}
\newcommand{\LEM}{\ll_{_{\mathrm L}}}
\newcommand{\UEM}{\ll_{_{\mathrm U}}}

\newcommand{\EMS}{\sqsubseteq_{_{\mathrm EM}}}
\newcommand{\LEMS}{\sqsubseteq_{_{\mathrm L}}}
\newcommand{\UEMS}{\sqsubseteq_{_{\mathrm U}}}

\newcommand{\im}{\mathrm{Im}}

\newcommand{\Cl}{\mathrm{Cl}}
\newcommand{\reals}{{\mathbb R}}
\newcommand{\IR}{{\mathbf I}\,\!{\mathbb R}}

\newcommand{\UX}{{\mathbf U}\,\!{\mathit X}}

\newcommand{\Pfin}{{\mathcal P}_{\mathit{fin}}}

\newcommand{\rat}{{\mathbb Q}}

\newtheorem{Th}{Theorem}[section]
\newtheorem{theorem}[Th]{Theorem}
\newtheorem{proposition}[Th]{Proposition}     
\newtheorem{lemma}[Th]{Lemma}
\newtheorem{corollary}[Th]{Corollary}
\newtheorem{definition}[Th]{Definition} 
\newtheorem{example}[Th]{Example}

\begin{document}

\title{Compactness of the space of causal curves}

\author{Keye Martin \\ \\
{\small Department of Mathematics}\\
{\small Tulane University}\\
{\small New Orleans, LA 70118}\\
{\small \texttt{martin@math.tulane.edu} $\cdot$ \texttt{http://math.tulane.edu/$\tilde{\ }$martin}}
}
\date{}
\maketitle

\begin{abstract} We prove
that the space of causal curves between
compact subsets of a separable globally hyperbolic
poset is itself compact in the Vietoris topology.
Although this result implies the usual
result in general relativity, its proof does not require
the use of geometry or differentiable structure.
\end{abstract}

\section{Introduction}

An important result in general relativity is
that the space of causal curves between two
compact sets in a globally hyperbolic spacetime is itself compact 
in the Vietoris (and hence upper) topology.
This result plays an important role in the proofs of
certain singularity theorems~\cite{wald},
in establishing the existence of maximum length
geodesics~\cite{he}, and in the proof
of certain positive mass theorems~\cite{penrose:mass}.

Recently~\cite{gr} it was shown that a globally
hyperbolic spacetime $\mathcal{M}$ with its causality relation
forms a special type of bicontinuous poset 
(i.e., a \em globally hyperbolic poset \em)
which has a 
canonical representation by a \em domain \em $\mathbf{I}\mathcal{M}$.  
This implies
for example that a globally hyperbolic spacetime
can be order theoretically reconstructed from only
a countable dense set of events and timelike causality.
It also suggests that the natural way to topologize
the space of causal curves in general relativity
is with the Vietoris topology. The reason for this is
as follows.

First,
there is a homeomorphism $\mathcal{M}\simeq\max(\mathbf{I}\mathcal{M})$
between the manifold topology and the relative \em Scott \em
topology on $\max(\mathbf{I}\mathcal{M})$. Thus,
\em events \em are maximal elements in $\mathbf{I}\mathcal{M}$.
Next, given any $\omega$-continuous domain $D$ with $\max(D)$ metrizable,
there is a `higher dimensional' domain $\mathbf{C}D$ called
the \em convex powerdomain \em which admits an injection
\[\Pcom(\max(D))\rightarrow\max(\convex D)\]
of the nonempty compact subsets of $\max(D)$ (in its relative
Scott topology) into the maximal elements of $\convex(D)$.
In particular, \em causal curves \em in a
globally hyperbolic spacetime are maximal elements 
in $\convex(\mathbf{I}\mathcal{M})$. Just as the
relative Scott topology on $\max(\mathbf{I}\mathcal{M})$
gives the manifold topology, the relative Scott topology
on $\max(\convex(\mathbf{I}\mathcal{M}))$, when restricted to
the causal curves, is exactly the Vietoris topology!
When coupled with the fact that
compactness in the Vietoris topology
implies compactness in the upper topology,
the use of the Vietoris topology
in~\cite{sorkin} is definitely natural,
and probably aesthetically necessary.

So as stated above we prove
that the space of causal curves between 
compact sets in a globally hyperbolic poset
is compact in the Vietoris topology. The
fact that the proof is entirely order theoretic
seems to provide evidence that globally hyperbolic posets
provide an abstract formulation of the physical
notion causality which is independent of geometry
and differentiable structure.

\section{Domains}

A \em poset \em is a partially ordered set, i.e., a set together
with a reflexive, antisymmetric and transitive relation.

\begin{definition}\em Let $(P,\sqsubseteq)$ be a partially ordered set.
  A nonempty subset $S\subseteq P$ is \em directed \em if $(\forall
  x,y\in S)(\exists z\in S)\:x,y\sqsubseteq z$. The \em supremum \em
  of $S\subseteq P$ is the least of all its upper bounds
  provided it exists. This is written $\bigsqcup S$.
\end{definition}
These ideas have duals
that will be important to us: A nonempty $S\subseteq P$ is \em filtered \em if $(\forall
  x,y\in S)(\exists z\in S)\:z\sqsubseteq x,y$. The \em infimum \em $\bigwedge S$
  of $S\subseteq P$ is the greatest of all its lower bounds
  provided it exists.

\begin{definition}\em For a subset $X$ of a poset $P$, set
\[\uparrow\!\!X:=\{y\in P:(\exists x\in X)\,x\sqsubseteq y\}\ \ \&\
\downarrow\!\!X:=\{y\in P:(\exists x\in X)\,y\sqsubseteq x\}.\] We
write $\uparrow\!x=\,\uparrow\!\{x\}$ and
$\downarrow\!x=\,\downarrow\!\{x\}$ for elements $x\in X$.
\end{definition}

A partial order allows for the derivation of
several intrinsically defined topologies. 
Here is our first example.

\begin{definition}\em  A subset $U$ of a poset $P$ is \em Scott open \em if
\begin{enumerate}
\item[(i)] $U$ is an upper set: $x\in U\ \&\ x\sqsubseteq y\Rightarrow
  y\in U$, and \item[(ii)] $U$ is inaccessible by directed suprema:
  For every directed $S\subseteq P$ with a supremum,
\[\bigsqcup S\in U\Rightarrow S\cap U\neq\emptyset.\]
\end{enumerate}
The collection of all Scott open sets on $P$ is called the \em Scott
topology. \em
\end{definition}

\begin{definition}\em
A \em dcpo \em is a poset in which every directed subset has a
  supremum. The \em least element \em in a poset,
  when it exists, is the unique element $\bot$ with $\bot\sqsubseteq x$
for all $x$.
\end{definition}

The set of \em maximal elements \em in a dcpo $D$ is
\[\max(D):=\{x\in D :\ \uparrow\!\!x=\{x\}\}.\]
Each element in a dcpo has a maximal
element above it.

\begin{definition}\em
  For elements $x,y$ of a poset, write $x\ll y$ iff for all directed
  sets $S$ with a supremum,
\[y\sqsubseteq\bigsqcup S\Rightarrow (\exists s\in S)\:x\sqsubseteq s.\]
We set $\dda x=\{a\in D:a\ll x\}$ and $\dua x=\{a\in D:x\ll a\}$.
\end{definition}
For the symbol ``$\ll$,'' read ``approximates.'' 

\begin{definition}\em
  A \em basis \em for a poset $D$ is a subset $B$ such that $B\cap\dda x$
  contains a directed set with supremum $x$ for all $x\in D$.  A poset is
  \em continuous \em if it has a basis. A poset is $\omega$-\em continuous \em
  if it has a countable basis.
\end{definition}

Continuous posets have an important property, they are \em interpolative. \em

\begin{proposition} If $x\ll y$ in a continuous poset $P$, then
there is $z\in P$ with $x\ll z\ll y$.
\end{proposition}

This enables a clear description of the Scott topology,

\begin{theorem}
  The collection $\{\dua x:x\in D\}$ is a basis for the Scott topology
  on a continuous poset.
\end{theorem}

And also helps us give a clear definition of the \em Lawson topology. \em

\begin{definition}\em The \em Lawson topology \em on a continuous poset $P$
has as a basis all sets of the form $\dua x\setminus\!\!\uparrow\!\!F$,
for $F\subseteq P$ finite.
\end{definition}

\begin{definition}\em 
A \em continuous dcpo \em is a continuous poset which is also a dcpo.
A \em domain \em is a continuous dcpo.
\end{definition}
We now consider some examples 
that illustrate the basic ideas. 

\begin{example}\em Let $X$ be a locally compact Hausdorff space. Its \em upper space \em
\[\UX=\{\emptyset\neq K\subseteq X:K\mbox{ is compact}\}\]
ordered under reverse inclusion
\[A\sqsubseteq B\Leftrightarrow B\subseteq A\]
is a continuous dcpo: 
\begin{itemize}
\item For directed $S\subseteq\UX$, $\bigsqcup S=\bigcap S.$ 
\item For all $K,L\in \UX$, $K\ll L\Leftrightarrow L\subseteq\mbox{int}(K)$.
\item $\UX$ is $\omega$-continuous iff $X$ has a countable basis.
\end{itemize}
It is interesting here that the space $X$ can be recovered
from $\UX$ in a purely order theoretic manner:
\[X\simeq\max(\UX)=\{\{x\}:x\in X\}\]
where $\max(\UX)$ carries the relative Scott topology it 
inherits as a subset of $\UX.$ Several constructions
of this type are known.
\end{example}
The next example is due to Scott\cite{scott:outline};
it is good to understand it in detail, especially 
since globally hyperbolic spacetimes admit a completely
analogous construction.
\begin{example}\em  The collection of compact intervals of the real line
\[\IR=\{[a,b]:a,b\in\reals\ \&\ a\leq b\}\]
ordered under reverse inclusion
\[[a,b]\sqsubseteq[c,d]\Leftrightarrow[c,d]\subseteq[a,b]\]
is an $\omega$-continuous dcpo:
\begin{itemize}
\item For directed $S\subseteq\IR$, $\bigsqcup S=\bigcap S$, 
\item
  $I\ll J\Leftrightarrow J\subseteq\mbox{int}(I)$, and \item
  $\{[p,q]:p,q\in\rat\ \&\ p\leq q\}$ is a countable basis for $\IR$.
\end{itemize}
The domain $\IR$ is called the \em interval domain.\em
\end{example}
We also have $\max(\IR)\simeq\reals$ in the Scott topology. 
The reason for this is that a basic Scott open set in $\IR$ has
the form
\[\dua[a,b]=\{x\in\IR:x\subseteq(a,b)\}\]
so when the Scott topology is restricted to $\max(\IR)$,
we get open sets of the form
\[\dua[a,b]\cap\max(\IR)=\{[x]:x\in(a,b)\}\simeq(a,b)\]
which is just the Euclidean topology on $\reals$.

\section{Globally hyperbolic posets}

In this section we consider a very special
example of a domain. But first:

\begin{definition}\em A continuous poset $P$ is \em bicontinuous \em if
\begin{itemize}
\item For all $x,y\in P$, $x\ll y$ iff for all filtered $S\subseteq P$
with an infimum,
\[\bigwedge S\sqsubseteq x\Rightarrow (\exists s\in S)\,s\sqsubseteq y, \]
and
\item For each $x\in P$, the set $\dua x$ is filtered with infimum $x$.
\end{itemize}
\end{definition}

\begin{example}\em $\reals$, $\rat$ are bicontinuous.
\end{example}

\begin{definition}\em On a bicontinuous poset $P$, sets of the form
\[(a,b):=\{x\in P:a\ll x\ll b\}\]
form a basis for a topology called \em the interval topology. \em
\end{definition}
That the open intervals form a basis
for a topology uses interpolation and bicontinuity. 
On a bicontinuous poset, the Lawson topology
is contained in the interval topology (causal simplicity),
the interval topology is Hausdorff (strong causality),
and $\leq$ is a closed subset of $P^2$. 
Globally hyperbolic posets provide
a special example of a domain.

\begin{definition}\em A poset $(X,\leq)$ is \em globally
hyperbolic \em if it is bicontinuous and if the sets
\[[a,b]:=\{x\in X:a\leq x\leq b\}\]
are compact in the interval topology.
\end{definition}

\begin{theorem} The closed intervals of a globally hyperbolic poset $X$ 
\[{\bf I}X:=\{[a,b]:a\leq b\ \&\ a,b\in X\}\]
ordered by reverse inclusion
\[[a,b]\sqsubseteq[c,d]\equiv [c,d]\subseteq [a,b] \]
form a continuous domain with
\[[a,b]\ll[c,d]\equiv a\ll c\ \&\ d\ll b.\]
$X$ has a countable basis iff ${\bf I}X$ is $\omega$-continuous. Finally,
\[\max({\bf I}X)\simeq X\]
where the set of maximal elements has the relative Scott
topology from ${\bf I}X$,
and $X$ has the interval topology.
\end{theorem}
In fact there is even an equivalence between
globally hyperbolic posets and certain
types of domains~\cite{gr}. We briefly now recall the relevance
of these ideas to general relativity;
more detailed definitions are in~\cite{gr}.

\begin{definition}\em A \em spacetime \em is a real four-dimensional
smooth manifold $\mathcal{M}$ with a Lorentz metric $g_{ab}$.
\end{definition}

Let $(\mathcal{M},g_{ab})$ be a time orientable spacetime.
Let $\Pi^+_\leq$ denote the future directed
causal curves, and $\Pi^+_{<}$ denote
the future directed time-like curves.

\begin{definition}\em For $p\in \mathcal{M}$, 
\[I^+(p):=\{q\in\mathcal{M}:(\exists\pi\in\Pi^+_{<})\,\pi(0)=p, \pi(1)=q\}\]
and
\[J^+(p):=\{q\in\mathcal{M}:(\exists\pi\in\Pi^+_{\leq})\,\pi(0)=p, \pi(1)=q\}\]
Similarly, we define $I^-(p)$ and $J^-(p)$.
\end{definition}
We write the relation $J^+$ as
\[p\leq q\equiv q\in J^+(p).\]

\begin{definition}\em A spacetime $\mathcal{M}$ is \em globally hyperbolic \em
if it is strongly causal and 
if $\uparrow\!\!a\ \cap\downarrow\!\!b$ is compact in the manifold topology,
for all $a,b\in\mathcal{M}$.
\end{definition}

\begin{theorem} If $(\mathcal{M},\leq)$ is globally hyperbolic, 
then $(\mathcal{M},\sqsubseteq)$ is a bicontinuous poset
with $\ll\ =I^+$ whose interval topology is the manifold topology.
\end{theorem}

Thus, a globally hyperbolic spacetime is a
globally hyperbolic poset when equipped with its
causality relation $\leq$.

\section{The space of causal curves}

A fundamental result in relativity is
that the space of causal curves between points is compact
on a globally hyperbolic spacetime. 
We use domains as an aid in proving this fact 
for any globally
hyperbolic poset (they are not necessary though). 
One advantage to involving
domains in the picture is that the Vietoris topology on causal
curves arises as the natural counterpart
to the manifold topology on events, 
so we can understand that its use
in~\cite{sorkin} is very natural.

In addition, other results on the compactness of the
space of causal curves,
such as those used to establish the existence
of maximum length geodesics, are easily derivable from
this one result. In fact, the length function
is continuous with respect to the Vietoris
topology as well, so it is possible to make a very
strong case that the Vietoris topology
is more natural than the one normally used (the ``upper'' topology).

\begin{definition}\em Let $D$ be a continuous dcpo. For subsets $A,B\subseteq D$, we define
relations
\begin{itemize}
\item $A\LEM B\Leftrightarrow(\forall a\in A)(\exists b\in B)\:a\ll b$
\item $A\UEM B\Leftrightarrow(\forall b\in B)(\exists a\in A)\:a\ll b$
\item $A\EM B\Leftrightarrow A\LEM B\ \&\ A\UEM B$
\end{itemize}
In the same way, we derive $\LEMS,\UEMS$ and $\EMS$ from the order $\sqsubseteq$ on $D.$
\em
\end{definition}

\begin{definition}\em The nonempty finite subsets of a space $X$ are denoted $\Pfin(X),$
while its nonempty compact subsets are written as $\Pcom(X).$
\em
\end{definition}

The set $\Pfin(D)$ together with $\EM$ is an abstract basis.
(See the appendix for more on abstract bases).

\begin{definition}\em The \em convex powerdomain \em $\convex D$ of a continuous dcpo $D$ is the ideal completion
of the abstract basis $(\Pfin(D),\EM)$.
\end{definition}

\begin{definition}\em
For a Scott compact $K\in\Pcom(D)$, we set
\[K^*=\{F\in\Pfin(D):F\EM K\}.\]
\end{definition}
Notice that this operation is also defined for elements of $\Pfin(D).$

\begin{proposition} For a continuous dcpo $D$, we have
\label{CDproperties}
\begin{enumerate}
\em\item[(i)]\em If $K\in\Pcom(D)$, then $K^*=\{F\in\Pfin(D):F\EM K\}\in\convex D$.
\em\item[(ii)]\em For ideals $I,J\in\convex D,$
\[I\ll J\Leftrightarrow (\exists F,G\in\Pfin(D))\:F\EM G\ \&\ I\subseteq F^*\subseteq G^*\subseteq J.\]
\em\item[(iii)]\em For $F\in\Pfin(D)$ and $I\in\convex D$, $F\in I\Leftrightarrow F^*\ll I$.
\em\item[(iv)]\em For $F,G\in\Pfin(D)$, $F^*\sqsubseteq G^*$ in $\convex D\Leftrightarrow F\EMS G$.
\end{enumerate}
\end{proposition}

\begin{definition}\em
\label{VietorisTopology}
The \em Vietoris hyperspace \em of a Hausdorff space $X$
is the set of all nonempty compact subsets $\Pcom(X)$ with the \em Vietoris topology: \em
It has a basis given by all sets of the form
\[\sigma(U_1,\cdots,U_n):=\{K\in\Pcom(X): K\subseteq\bigcup_{i=1}^n U_i\ \mbox{and}\ K\cap U_i\neq\emptyset, 1\leq i\leq n\},\]
where $U_i$ is a nonempty open subset of $X$, for each $1\leq i\leq n.$
\end{definition}

The next result is from~\cite{martin:fractals}:
\begin{theorem}
\label{MeasurementOnCD}
Let $D$ be an $\omega$-continuous dcpo with $X=\max(D)$ regular in
its relative Scott topology. Then
the correspondence
\[\Pcom(X)\stackrel{e}{\longrightarrow}\max(\convex D)::K\mapsto K^*\]
is a homeomorphism between the Vietoris hyperspace of $X$ and $\im(e)$ in
its relative Scott topology.
\end{theorem}

Thus, events in
spacetime are maximal elements in ${\bf I}X$,
causal curves are maximal elements in $\mathbf{C}({\bf I}X)$. The reason
we know this is:

\begin{proposition} A subset of a globally hyperbolic
spacetime $\mathcal{M}$ is the image of a causal curve iff it is the image
of a continuous monotone increasing $\pi:[0,1]\rightarrow\mathcal{M}$ iff
it is a compact connected linearly ordered subset of $(\mathcal{M},\sqsubseteq)$.
\end{proposition}
This suggests the following:
\begin{definition}\em Let $(X,\leq)$ be a globally
hyperbolic poset. A subset $\pi\subseteq X$ is
a \em causal curve \em if it is compact, connected and linearly ordered.
We define 
\[\pi(0):=\bot\ \ \mbox{and}\ \ \pi(1):=\top\]
where $\bot$ and $\top$ are the least 
and greatest elements of $\pi$. For $P,Q\subseteq X$,
\[C(P,Q):=\{\pi:\pi\mbox{ causal curve},\pi(0)\in P, \pi(1)\in Q\}\]
and call this \em the space of causal curves \em between $P$ and $Q$.
\end{definition}

Let $D$ be an $\omega$-continuous dcpo
with $\max(D)$ regular. For example, it could be 
$D=\mathbf{I}X$ for $X$ globally hyperbolic and separable.

\begin{lemma} If $\uparrow\!\!x\cap\max(D)$ is Scott compact in $D$,
then $\uparrow\!\!\{x\}^*\cap\im(e)$ is Scott compact in $\convex D$.
\end{lemma}
{\bf Proof}. Let $L:= (\uparrow\!\!x\cap\max(D))$. It is a compact metric
space.  If $K^*\in\ \uparrow\!\!\{x\}^*\cap\im(e)$, then
$\{x\}^*\sqsubseteq K^*$, which implies $K\subseteq L$.
Thus, $\uparrow\!\!\{x\}^*\cap\im(e)\simeq \Pcom(L)$ in the Vietoris
topology, which is also a compact metric space~\cite{engelking}.
\endproof\newline

For the next lemma, it is important to point out that
$X=\max(D)$ is metrizable and thus \em normal. \em

\begin{lemma} If $(k_i)$ is a convergent sequence in $\Pcom(X)$ with each
$k_i$ connected, then $\lim k_i$ is connected.
\end{lemma}
{\bf Proof}. Assume $k:=\lim k_i$ is disconnected. Then $k=a\cup b$ is
a disjoint union of closed sets. Since $k$ is compact, $a$ and $b$
are compact, so $a^*,b^*\in\im(e)$. By normality of $X$, there are disjoint
open sets $U,V\subseteq X$ with $a\subseteq U$, $b\subseteq V$.
Thus, there are finite sets $F, G\in\Pfin(D)$ with 
$a\subseteq\dua F\cap X\subseteq U$ and $b\subseteq \dua G\cap X\subseteq V$.
If any element of $F$ or $G$ is not way below
some element of $a$ or $b$, we can simply throw it out. Thus,
\[F^*\ll a^*\ \mbox{and}\ G^*\ll b^*\ \Rightarrow\ (F\cup G)^*\ll (a\cup b)^*=k^*\]
so for large enough $i$, we have $(F\cup G)^*\ll (k_i)^*$,
since $(k_i)^*\rightarrow k^*$ in the relative Scott topology.
The sets $\dua F\cap k_i\cap X$ and $\dua G\cap k_i\cap X$
are open, disjoint and their union is $k_i$. They are
nonempty because $F\cup G\EM k_i$. Then $k_i$ is disconnected.
\endproof\newline

In the next lemma, $D=\mathbf{I}X$ the domain
of spacetime intervals for a separable globally hyperbolic poset.

\begin{lemma} If $(k_i)$ is a convergent sequence in $\Pcom(X)$ with each
$k_i$ linearly ordered, then $\lim k_i$ is linearly ordered.
\end{lemma}
{\bf Proof}. Let $k:=\lim k_i$ and $p,q\in k$. There
are increasing sequences $(x_n)$ and $(y_n)$ in $\mathbf{I}X$
with $x_n\ll [p]$, $y_n\ll [q]$, $\bigsqcup x_n=[p]$ and $\bigsqcup y_n=[q]$.
By compactness of $k$, extend $\{x_n,y_n\}$
to a finite $F_n\supseteq\{x_n,y_n\}$ with $F_n^*\ll k^*$.
Then
\[ (\forall n)(\exists i_n)\,(k_{i_n})^*\in\dua F_n^*\]
Since $x_n,y_n\in F_n\EM k_{i_n}$, there are $p_n,q_n\in k_{i_n}$
with $x_n\ll [p_n]$, $y_n\ll [q_n]$.
Then $p_n\rightarrow p$ and $q_n\rightarrow q$. 
Because $k_{i_n}$ is linearly ordered,
we either have $p_n\leq q_n$ for an infinite number of $n$
or $q_n\leq p_n$ for an infinite number of $n$.
Because $\leq$ is closed, either $p\leq q$ or $q\leq p$,
so $k$ is linearly ordered.
\endproof\newline

In our final lemma, $(X,\leq)$ is a separable globally hyperbolic poset.

\begin{lemma} 
\label{lastone}
Let $(\pi_n)$ be a sequence of causal curves with $\pi_n\rightarrow\pi$
in the Vietoris topology on $\Pcom(X)$. If $(\pi_n(0))$ and $(\pi_n(1))$
are both convergent, then $\pi_n(0)\rightarrow\pi(0)$
and $\pi_n(1)\rightarrow\pi(1)$.
\end{lemma}
{\bf Proof}. First, we prove
that for every $r\in\pi$, there is a subsequence $\alpha_n=\pi_{k_n}$
such that $r_n\rightarrow r$ where $r_n\in\alpha_n$. Take an increasing
sequence $(x_n)$ in $\mathbf{I}X$ with $x_n\ll[r]$ and $\bigsqcup x_n=[r]$.
Extend $\{x_n\}$ to a finite set $F_n\supseteq\{x_n\}$ with 
$F_n\EM\pi$ and thus $F_n^*\ll\pi^*$. For each $n$, let
$k_n$ be the least integer such that $F_n^*\ll\pi_{k_n}^*$ and hence $F_n\EM \pi_{k_n}$.
Since $x_n\in F_n$, there is $r_n\in\pi_{k_n}$ with $x_n\ll r_n$.
But $\bigsqcup x_n=[r]$, $r_n\rightarrow r$,
so $\alpha_n:=\pi_{k_n}$ is the subsequence.

Next, we claim $a:=\lim \pi_n(0)$ and $b:=\lim \pi_n(1)$ belong to $\pi$.
If $a\not\in\pi$, then $\pi$ and $\{a\}$ are disjoint compact sets,
so by normality there are disjoint open sets $U,V\subseteq X$ with
$\pi\subseteq U$ and $a\in V$. But, 
since $\pi_n(0)\rightarrow a$, $\pi_n(0)\in V$ for all $n\geq L$,
for some $L$.
And, since $\pi_n\rightarrow\pi$ in the Vietoris topology,
$\pi_n\subseteq U$ for all $n\geq L$, for some $L$. Thus,
for $n$ sufficiently large, $\pi_n(0)\in U\cap V$,
contradicting $U\cap V=\emptyset$.
A similar argument shows $b\in\pi$.

Finally, we can prove the claim. Given $r\in\pi$,
take a subsequence $(\alpha_n)$ of $(\pi_n)$ with
$r_n\rightarrow r$ and $r_n\in\alpha_n$. Since
$(\alpha_n)$ is a subsequence of $(\pi_n)$,
each $\alpha_n$ is linearly ordered, so
$\alpha_n(0)\leq r_n\leq\alpha_n(1)$. Then
since $\leq$ is closed,
\[\lim_{n\rightarrow\infty}\alpha_n(0)=\lim_{n\rightarrow\infty}\pi_n(0)\leq r\leq
\lim_{n\rightarrow\infty}\alpha_n(1)=\lim_{n\rightarrow\infty}\pi_n(1).\]
But $\lim \pi_n(0)$ and $\lim \pi_n(1)$
both belong to $\pi$ and $\pi$ is \em linearly ordered, \em
so $\lim \pi_n(0)=\pi(0)$ and $\lim \pi_n(1)=\pi(1)$
since $r$ was \em arbitrary. \em
\endproof

\begin{theorem} If $(X,\leq)$ is a separable globally hyperbolic poset, 
then the space of causal curves $C(P,Q)$ is compact in the Vietoris topology
when $P,Q\subseteq X$ are compact.
\end{theorem}
{\bf Proof}. Let $(x_n)$ be a sequence in $C(P,Q)$. The endpoints 
$(x_n(0))$ have a convergent subsequence in $P$, so we pass
to a subsequence of $(x_n)$ called $(y_n)$ with $(y_n(0))$ convergent. Then
$(y_n(1))$ has a convergent subsequence in $Q$, so we pass
to a subsequence of $(y_n)$ called $(\pi_n)$ which
has the property that both $(\pi_n(0))$ and $(\pi_n(1))$
are convergent.

Define $a=\lim \pi_n(0)\in P$ and $b=\lim \pi_n(1)\in Q$. By definition,
we have $\pi_n(0)\leq\pi_n(1)$ and thus $a\leq b$ since $\leq$ is closed.
Then $[a,b]\in\mathbf{I}X$ so take some $u\ll[a,b]$ and
notice that $\dua u\cap\max(\mathbf{I}X)$ is an open set around $[a],[b]$.
Since $\pi_n(0)\rightarrow a$ and $\pi_n(1)\rightarrow b$,
we can assume $\pi_n(0),\pi_n(1)\in u$ for all $n$. But
this implies $\pi_n\subseteq u$ because $u$ is an interval!
By global hyperbolicity, 
$\uparrow\! u\cap\max(\mathbf{I}X)$ is compact,
so $\uparrow\! \{u\}^*\cap\im(e)$ is compact in $\convex(\mathbf{I}X)$.
Then $(\pi_n)$ has a convergent subsequence in $\Pcom(u)$
which we simply call $(\pi_n)$. Thus, $\pi_n\rightarrow\pi\in\Pcom(u)$.

By our previous work, $\pi$ is compact, connected and linearly
ordered as a subset of $(X,\leq)$, so $\pi$ is a causal curve.
We still need to prove $\pi\in C(P,Q)$. By
Lemma~\ref{lastone},
\[\lim \pi_{n}(0)=\pi(0)\ \&\ \lim \pi_{n}(1)=\pi(1)\]
but we know these limits are exactly $a\in P$ and $b\in Q$.
This proves $\pi\in C(P,Q)$. Since every sequence has
a convergent subsequence and since $C(P,Q)\subseteq\Pcom(X)$ has
a countable basis (inherited from $X$), this
proves $C(P,Q)$ is compact.
\endproof\newline

The separability is probably not required to prove
this result, but we don't feel like working with
nets anymore. The result above has been applied 
in the proof of a positive mass theorem~\cite{penrose:mass}. We can also use it to 
deduce an important corollary, which is often used to establish the existence of maximum length
geodesics on globally hyperbolic spacetimes (\cite{he}\cite{wald}).

\begin{definition}\em The \em upper topology \em on $\Pcom(X)$
has as a basis sets of the form
\[O(U):=\{K\in\Pcom(X):K\subseteq U\}\]
where $U\subseteq X$ is open.
\end{definition}

\begin{corollary} If $(X,\leq)$ is a globally hyperbolic poset, 
the space of causal curves $C(P,Q)$ is compact in the upper topology whenever $P,Q\subseteq X$ are compact.
\end{corollary}
{\bf Proof}. The upper
topology is contained in the Vietoris topology.
\endproof\newline

A spacetime is globally hyperbolic iff either
of the space of causal curves is compact;
the nontrivial direction is the one we have abstracted
here to the level of posets. For separable globally
hyperbolic posets which `mimic' spacetime we
can also prove the following equivalence:

\begin{proposition} Let $(X,\leq)$ be a separable
locally compact bicontinuous poset such that $x\leq y$
iff there is a continuous monotone increasing ${\pi:[0,1]\rightarrow X}$
with $\pi(0)=x$ and $\pi(1)=y$. The following are
equivalent:
\begin{enumerate}
\em\item[(i)]\em The poset $(X,\leq)$ is globally hyperbolic.
\em\item[(ii)]\em The space $C(P,Q)$ is compact in the Vietoris
topology when $P,Q\subseteq X$ are compact.
\em\item[(iii)]\em The space $C(P,Q)$ is compact in the upper
topology when $P,Q\subseteq X$ are compact.
\end{enumerate}
\end{proposition}
{\bf Proof}. To prove (iii) $\Rightarrow$ (i), let $r_n\in[p,q]$
be a sequence. Then there is $\alpha_n:[0,1/2]\rightarrow X$ from
$p$ to $r_n$ and $\beta_n:[1/2,1]\rightarrow X$ from $r_n$ to $q$.
We can paste them together to get $\pi_n:[0,1]\rightarrow X$
from $p$ to $q$. Because $\pi_n$ is continuous and monotone,
its image $\im(\pi_n)$ is a causal curve. Since $C(\{p\},\{q\})$
is compact in the upper topology, $(\im(\pi_n))$ has a 
subsequence named $(\alpha_n)$ with $\alpha_n\rightarrow\alpha$.
By local compactness, there is an open set $U\subseteq X$ with $\Cl(U)$ compact and 
$\alpha\subseteq U$. Then there is an integer
$N$ such that $r_{k_n}\in\alpha_n\subseteq\Cl(U)$ for all $n\geq N$.
By compactness, the sequence $(r_{k_n})$ has a
convergent subsequence with limit $r$.
But $r_{k_n}\in\alpha_n\subseteq[p,q]$
and $[p,q]$ is closed as the intersection of Lawson closed sets, 
so $r\in[p,q]$.

This proves $(r_n)$ has a convergent subsequence in $[p,q]$
which by second countability proves compactness. \endproof\newline

The property required of $(X,\leq)$ is
interesting because it is one that domains
in quantum mechanics also satisfy~\cite{martin:wheel}.
In closing this section, the fact that the compactness
result can be proven for globally hyperbolic
posets provides evidence that we have
identified a useful order theoretic formulation
of causality that does not use
geometry or differentiable structure.

\section{Conclusion}

Sorkin and Woolgars in~\cite{sorkin} recast
the tools of ``global causal analysis'' using
only topology and order. Here we have gone a
step further, showing that only order is necessary,
the topology is implicitly described by the order
at an abstract level. Our results apply
even to posets that are not
spacetimes, demonstrating that in no way
can our proof depend on geometry. However, we owe
a great debt to the paper~\cite{sorkin},
which in our mind very clearly pointed the way to the
proof of our main result.

\section*{Appendix: Abstract bases}

A useful technique for constructing domains is to take the \em ideal completion \em of an \em abstract basis. \em

\begin{definition}\em An \em abstract basis \em is given by a set $B$ together with a transitive relation $<$ on $B$
which is \em interpolative, \em that is, 
\[M < x\Rightarrow (\,\exists\,y\in B\,)\:M < y < x\]
for all $x\in B$ and all finite subsets $M$ of $B$.
\end{definition}

Notice the meaning of $M<x$: It means $y<x$ for all $y\in M$. 
Abstract bases are covered in~\cite{abramsky:domain}, which
is where one finds the following. 

\begin{definition}\em An \em ideal \em in $(B,<)$ is a nonempty subset $I$ of $B$ such that
\begin{enumerate}
\item[(i)] $I$ is a lower set: $(\,\forall\, x\in B\,)(\,\forall\, y\in I\,)\:x<y\Rightarrow x\in I.$
\item[(ii)] $I$ is directed: $(\,\forall\, x,y\in I\,)(\,\exists\, z\in I\,)\:x,y<z.$
\end{enumerate}
The collection of ideals of an abstract basis $(B,<)$ ordered under inclusion is a partially ordered set
called the \em ideal completion \em of $B.$ We denote this poset by $\bar{B}.$
\end{definition}

The set $\{y\in B:y<x\}$ for $x\in B$ is an ideal which leads to a natural mapping from $B$ into $\overline{B}$,
given by $i(x)=\{y\in B:y<x\}.$

\begin{proposition} If $(B,<)$ is an abstract basis, then
\begin{enumerate}
\em\item[(i)]\em Its ideal completion $\bar{B}$ is a dcpo.
\em\item[(ii)]\em For $I,J\in\bar{B}$,
\[I\ll J\Leftrightarrow (\,\exists\,x,y\in B\,)\:x< y\ \&\ I\subseteq i(x)\subseteq i(y)\subseteq J.\] 
\em\item[(iii)]\em $\bar{B}$ is a continuous dcpo with basis $i(B).$
\end{enumerate}
\end{proposition}

If one takes any basis $B$ of a domain $D$ and restricts the approximation relation $\ll$ on $D$ to $B$,
they are left with an abstract basis $(B,\ll)$ whose ideal completion is $D$. Thus, all domains arise
as the ideal completion of an abstract basis.

\end{document}